\documentclass[a4paper, 10pt, conference]{IEEEtran}

\usepackage{graphics} % for pdf, bitmapped graphics files
\usepackage{epsfig} % for postscript graphics files
\usepackage{mathptmx} % assumes new font selection scheme installed
\usepackage{times} % assumes new font selection scheme installed
\usepackage{amsmath} % assumes amsmath package installed
\usepackage{amssymb}  % assumes amsmath package installed
\usepackage{subfigure}
\usepackage{multirow}
\usepackage{url}
\usepackage{epstopdf}
\usepackage{color}
\usepackage{natbib}
\usepackage{tikz}
\usepackage{gnuplot-lua-tikz}
\usepackage{tikz-uml}

\author{Najia Bouha$^{\star \star}$ , Gildas Morvan$^\star$, Hassane Aboua\"issa$^\star$, Yoann Kubera$^\star$ \\$^\star$ Univ. Lille Nord France, 59000 Lille, France,\\
U-Artois, LGI2A (EA 3926), \\ Technoparc Futura, 62400 B\'ethune, France\\ Email: (hassane.abouaissa, gildas.morvan, yoann.kubera)@univ-artois.fr\\
$\star \star$ Universit\'e Ibn Zohr, Facult\'e des Sciences,\\ Agadir, Morocco, 
\\
Email: najia.bouha@gmail.com}

\title{\textbf{\textsc{A First Step Towards Dynamic Hybrid Traffic Modeling}}}
\begin{document}
\maketitle

% Suppress page numbering 
\thispagestyle{empty}\pagestyle{empty}

%% Do not use the following keywords and abstract environment, use \section instead to get desired layout
%%\begin{keywords}
%%Agent-based modelling; Genetic algorithms
%%\end{keywords}
%%
%%\begin{abstract}
%%Abstract goes here. This is an example of using the ECMS \LaTeX\ class.
%%\end{abstract}

%% Notes about sections: Use star * to avoid numbering of sections

\section*{\textbf{KEYWORDS}}

Agent-based modeling; Multi-level modeling; Intelligent transportation systems; Simulation; Traffic flow
\section*{\textbf{ABSTRACT}}
{Hybrid traffic modeling and simulation provide an important way to represent and evaluate large-scale traffic networks at different levels of details. The first level, called ``\textit{microscopic}" allows the description of individual vehicles and their interactions as well as the study of driver's individual behavior. The second, based on the analogy with fluidic dynamic, is the ``\textit{macroscopic}" one and provides an efficient way to represent traffic flow behavior in large traffic infrastructures, using three aggregated variables: traffic density, mean speed and traffic volume. An intermediate level called ``\textit{mesoscopic}" considers a group of vehicles sharing common properties such as a same origin and destination. The work conducted in this paper presents a first step allowing simulation of wide area traffic network on the basis of \textit{dynamic hybrid modeling}, where the representation associated to a network section can change at runtime. The proposed approach is implemented in a simulation platform, called \textsc{Jam-free}.}
 
\section*{\textbf{INTRODUCTION}}
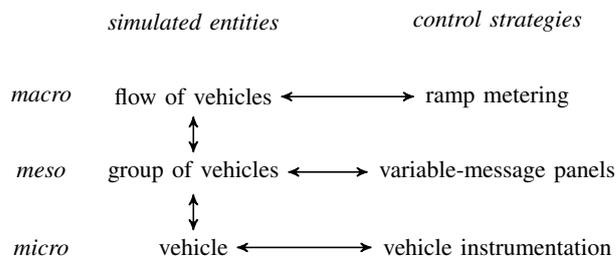
\begin{figure}[h]
\begin{center}
{\small \begin{tikzpicture}[>=stealth',shorten >=1pt,auto,semithick]
\node (micro) {\textit{micro}} ;
\node[above of=micro,node distance=1cm]  (meso) {\textit{meso}} ;
\node[above of=meso,node distance=1cm]   (macro) {\textit{macro}} ;

\node[right of=micro,node distance=2cm]  (vehicle) {vehicle} ;
\node[right of=meso,node distance=2cm]  (group) {group of vehicles} ;
\node[right of=macro,node distance=2cm]  (flow) {flow of vehicles} ;
\node[right of=micro,node distance=6cm]  (instrumentation) {vehicle instrumentation} ;
\node[right of=meso,node distance=6cm]  (panels) {variable-message panels} ;
\node[right of=macro,node distance=6cm]  (metering) {ramp metering} ;

\node[above of=flow,node distance=1cm] {\textit{simulated entities}} ;
\node[above of=metering,node distance=1cm] {\textit{control strategies}} ;

\draw[<->] (vehicle) -- node {} (group)   ;
\draw[<->] (group) -- node {} (flow)   ;
\draw[<->] (vehicle) -- node {} (instrumentation)   ;
\draw[<->] (group) -- node {} (panels)   ;
\draw[<->] (flow) -- node {} (metering)   ;
\end{tikzpicture} }
\caption{Hybrid traffic simulation and control approach}
\label{mlmtraffic}
\end{center}
\end{figure}

{Severe congestion is a daily problem which leads to a continuously growth of direct and indirect cost. Traffic congestion which can be recurrent typical to rush hours or non recurrent due to accidents, works,~\dots~represents a major preoccupation of many transportation institutions and practitioners, calls for an efficient and intelligent dynamic management.  Accordingly, several  works were undertaken to study the traffic phenomena, to implement effective strategies for an optimal use of the existing infrastructure and to minimize the congestion effects as well as a high quality of service. Nevertheless, the implementation of traffic measurement and control algorithms calls for a deep understanding of the traffic phenomena. In this context, traffic flow modeling and simulation play an important role and constitute efficient tools to perform tasks such as traffic prediction and monitoring, traffic control and forecasting, the repercussion of the construction of new parts on infrastructure onto the global behavior of the traffic flow,~\dots~According to the defined objective, several models have been developed and can be classified into microscopic, mesoscopic and macroscopic models. However, to simulate large-scale road networks, it can be interesting to integrate different representations in the same framework which leads to the so-called ``hybrid modeling" as shown on fig.~\ref{mlmtraffic}. Note that the concept of hybrid modeling has different meanings according to the studied domain. Here, hybrid modeling means the coupling of different models.  

In this paper, the first step is devoted to the integration of microscopic and macroscopic models into a single framework. The concept of hybrid modeling has been developed by several authors. Hence, some existing hybrid micro-macro traffic models  are shown in Table~\ref{hybridmodels}.}

\begin{table}[h]
\begin{center}
\begin{tabular}{ccc}
\hline
model&micro model&macro model\\
\hline
\citet{Magne:2000}&SITRA-B+&SIMRES\\
\citet{Poschinger:2002}&IDM&Payne\\
\citet{Bourrel:2003}&\multirow{2}{*}{optimal velocity}&LWR\\
\citet{Mammar:2006}&&ARZ\\
\citet{Espie:2006}&ARCHISM&SSMT\\
\citet{El-hmam:2006c}&generic ABM&LWR, ARZ, Payne\\
\citet{Joueiai:2013}&IDM&LWR\\
\hline
\end{tabular}
\caption{Static micro-macro traffic flow models proposed in the literature} \label{hybridmodels}
\end{center}
\end{table}%

The models presented in Table~\ref{hybridmodels} share the same limitation: connections between levels are fixed \textit{a priori} and cannot be changed at runtime. Therefore, to be able to observe some emerging phenomena such as congestion formation or to find the exact location of a jam in a large macro section, a dynamic hybrid modeling approach is needed.

The work presented in this paper is devoted to overcome these shortcomings and proposes the first step towards the development and implementation of dynamic hybrid models. Such a new approach provides an efficient way to change the level of representation dynamically. As a result of these development, a platform called \textsc{Jam-free} has been implemented.

In the following sections, we identify the main issues related to dynamic hybrid traffic modeling and then propose some generic solutions to them. Then, the \textsc{Jam-free} simulator is presented, as well as some simulation results. Finally, we conclude and suggest some research perspectives.

\section*{Dynamic hybrid traffic modeling: motivations and existing solutions}\label{dhm}

\subsection*{Motivations}

Traffic simulation is generally used to:
\begin{enumerate}
\item simulate the road traffic flow on very large and complex infrastructures including both city and highways while
\item being able to test the effects of  different  traffic signs, for example, or {traffic assignment and dynamic routing} strategies.  {It provides also a simple way to evaluate the impact of dynamic traffic control (ramp-metering, dynamic speed limit,~\dots) on the whole behavior of the traffic},
\item assess their precise influence on the traffic flow and 
\item understand {how and} why traffic perturbations {leading to appearance of}  {congestions}  do occur.
\end{enumerate}

The first goal is easily achieved using macroscopic simulation models at the expense of the fourth goal. Conversely, the fourth goal is easily {carried-out} using microscopic simulation models at the expense of the first goal. To deal with this paradox, and benefit from both approaches, a dynamic hybrid model can be used.

The advantages of this hybrid approach include the ability:
\begin{itemize}
\item To obtain both quantitative and qualitative information about the road traffic, using respectively macroscopic and microscopic representations in the same simulation {using the clusters principle}.
\item to switch between these {different levels of representation} locally depending:
\begin{itemize}
\item on the simulation needs; for instance understanding the source of a traffic {congestion},
\item on the computation constraints; for instance managing the CPU load,
\end{itemize}
\item to experiment both macroscopic and microscopic {dynamic routing strategies, \textit{i.e.} road load balancing strategies, and others dynamic trafic control}.
\end{itemize}

\subsection*{Use cases examples}

In order to assess the relevance of our  approach for dynamic hybrid simulation, use cases demonstrating the limits of the existing solutions depicted in Table~\ref{hybridmodels} are presented in this section.

\subsubsection*{Reduce the CPU load during peak hours}

\begin{figure*}[htbp]
	\includegraphics[width=\linewidth]{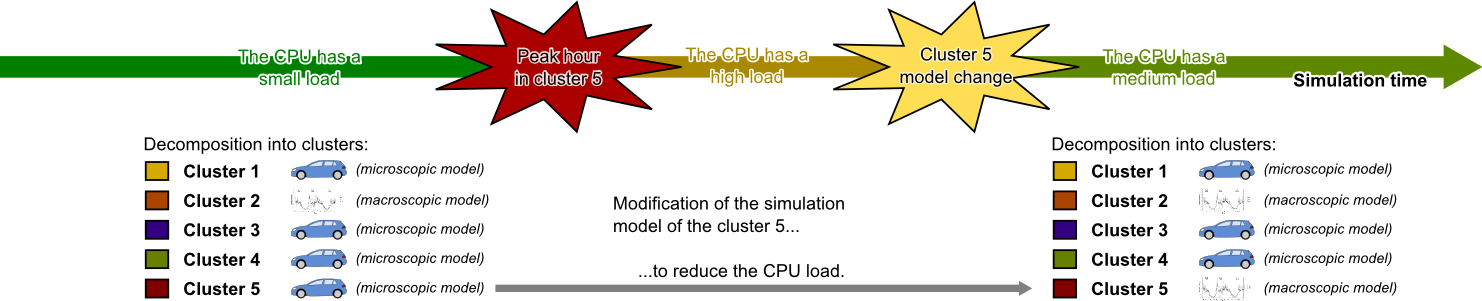}
	\caption{Illustration of the use case "Reduce the CPU load during peak hours".}
	\label{img:usecase CPU reduce peak hours}
\end{figure*}

The hybrid model can change the traffic representation of a portion of the road network dynamically. This feature can be used to switch from a microscopic representation to a macroscopic  {one}, when the CPU is overused in order to reduce its load. Such situations include cases where the number of simulated vehicles becomes too high to be managed satisfyingly by the CPU (for instance during peak hours). This case is illustrated in figure~\ref{img:usecase CPU reduce peak hours}.

\subsubsection*{Balance the CPU load between clusters}

The hybrid model can change the traffic representation of a portion of the road network dynamically. This feature can be used to switch from:
\begin{itemize}
	\item a microscopic representation to a macroscopic one when the CPU is overused in order to reduce its load,
	\item {inversely (from macroscopic to microscopic), in order to detect the reason for the occurrence of traffic congestion}.
	\end{itemize}
These features can be used jointly to balance the  {CPU} load between two clusters.

\begin{figure*}[htbp]
	\includegraphics[width=\linewidth]{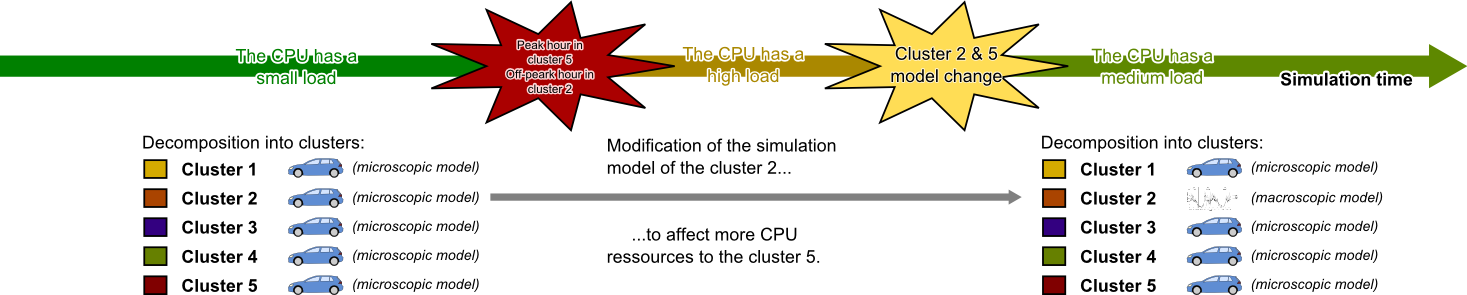}
	\caption{Illustration of the use case "Balance the CPU load between clusters".}
	\label{img:usecase cpu load balance 2}
\end{figure*}

Such situations include cases where a traffic jam (or congestion) appears in a cluster using a macroscopic model and where the number of simulated vehicles in another cluster is significantly reduced (for instance during off-peak hours).
Such situations also include cases where a traffic jam appears in a cluster using a microscopic model and where the number of simulated vehicles in another cluster is significantly reduced (for instance during fluidic period of the flow) as illustrated in figure~\ref{img:usecase cpu load balance 2}.

\subsubsection*{Find out the reason of {the occurence of the } traffic jam}

\begin{figure*}[htbp]
	\includegraphics[width=\linewidth]{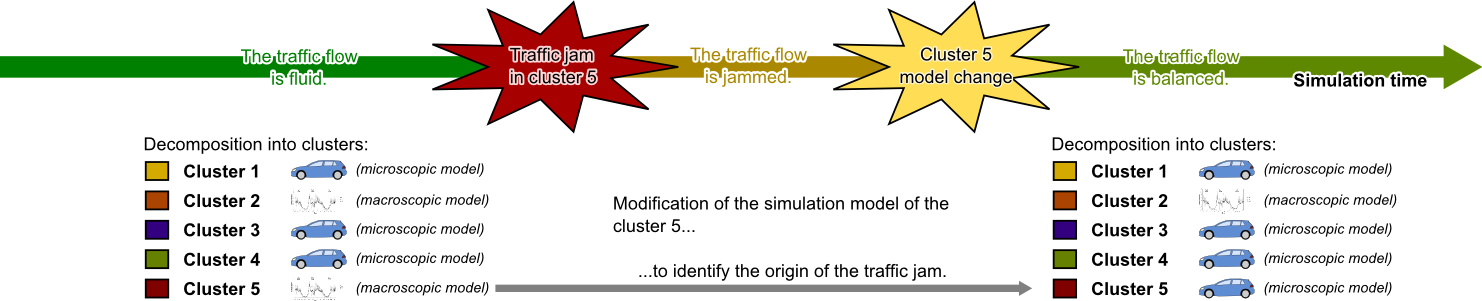}
	\caption{Illustration of the use case "Find out the reason of a traffic jam".}
	\label{img:usecase jams reasons}
\end{figure*}

{Figure~\ref{img:usecase jams reasons} shows an example of the occurence of traffic congestions. It demonstrates the relevance of dynamic hybrid simulations to switch between two levels of representation in order to find out the  causes of the appearance of these phenomena. } 

\subsubsection*{Follow moving macroscopic phenomena}

\begin{figure*}[htbp]
	\includegraphics[width=\linewidth]{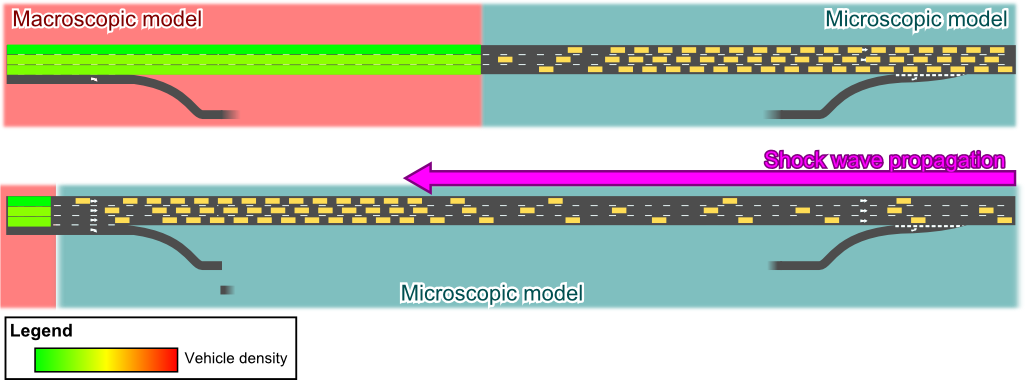}
	\caption{Dynamic hybrid model of a shockwave phenomenon.}
	\label{img:usecase shockwave dynamic}
\end{figure*}

The static division of the road network into clusters that can be found in most {of the} existing approaches causes precision losses. Indeed, a traffic flow simulator has to be able to reproduce macroscopic phenomena such as shockwaves, capacity drop,~\dots
Yet, the static decomposition found in static hybrid models will cause eventually a loss {of} quantitative {information}. To avoid this issue, a dynamic hybrid model gives the ability to move, resize, split and merge the clusters of the road network (see figure~\ref{img:usecase shockwave dynamic}).

\subsection*{Limits of the existing solutions}

To the best of our knowledge, the only other work describing a dynamic hybrid model is~\citet{Sewall:2011}. However, authors focus on the visualization of data rather than accurate simulation of traffic flow: zooming in a part of the network triggers a micro representation on this area. Thus, there is no bi-directional information exchange between the micro and macro representations.

\section*{Dynamic hybrid traffic modeling: issues and proposed solutions}

\subsection*{Main issues}

First, we identify the main issues to address {in order} to simulate dynamic hybrid traffic models accurately.
\begin{enumerate}
\item How can the simulated network be structured to switch between different models at runtime?
\item How to switch between different models (particularly from macro to micro)?
\item How the temporalities of the different interacting models can be managed to avoid bias?
\end{enumerate}
In the next section, we present possible generic solutions to these issues.

\subsection*{Proposed solutions}

\subsubsection*{Dynamic decomposition of the network into autonomous clusters}

The network can be dynamically decomposed into autonomous clusters. Each cluster operates autonomously and simulates the traffic flow using a model of its own. However, cutting the road network is not arbitrary, to avoid incoherent situation like having an area of a few square meters. To ensure the integrity of the simulation, a minimum cut is deducted from sensors: the entry and exit points of a cluster are necessarily sensors. Thus, the road network can be viewed as a directed graph of interconnected sensors. Each arc of the graph represents a path joining two sensors directly, without going through an intermediate sensor.

A cluster is then characterized by input and output sensors and eventually sensors situated inside the cluster. The minimum cutting of the road network is then defined by:
\begin{itemize}
	\item a cluster representing the outside of the simulated network,
	\item clusters representing minimal subsets in the road network; a cluster is minimal if there is no sensor inside the cluster.
\end{itemize}

\begin{figure*}[ht]
	\centering
	\includegraphics[width=\linewidth]{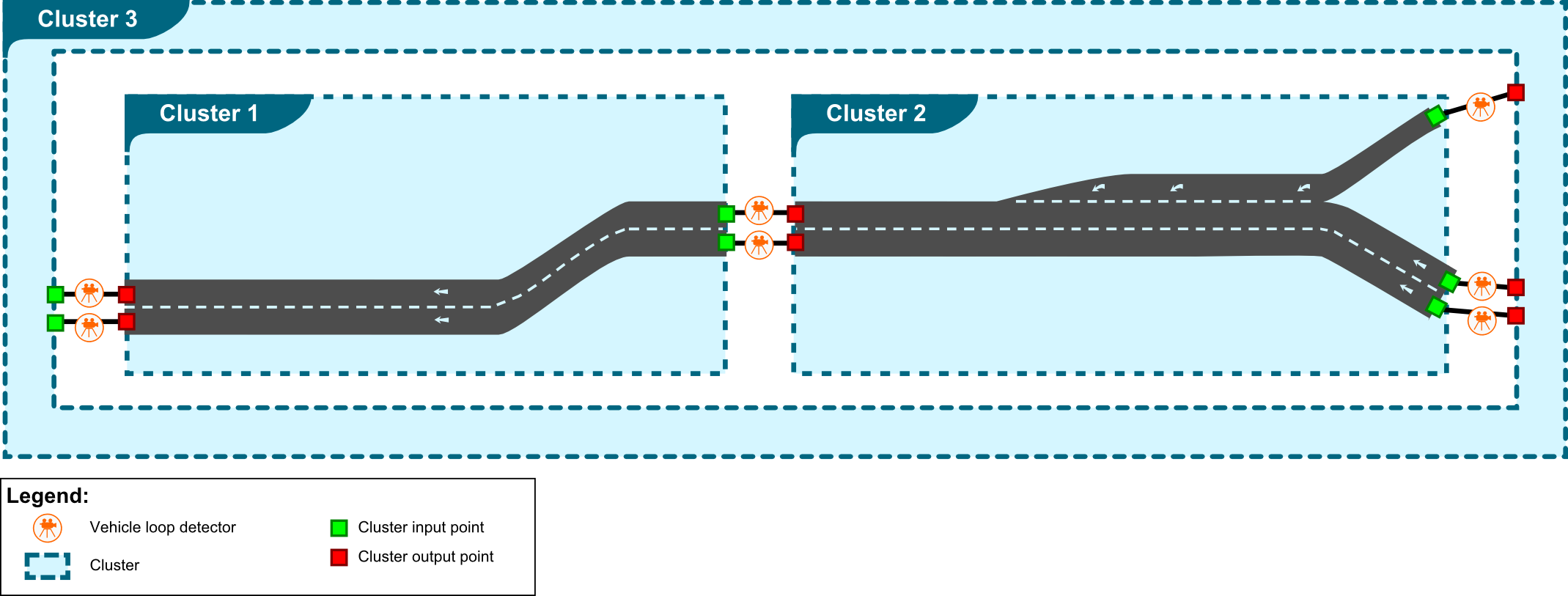}
	\caption{Minimal cutting of a network}
	\label{img:minimal clusters}
\end{figure*}

Clusters of a simulation are necessarily disjoint combination of minimum contiguous clusters. An example of minimal cutting is depicted in figure~\ref{img:minimal clusters}. Each cluster transmits mean speed and flow information from its bounding sensors to the upstream and downstream clusters. In the case of a macro cluster, this information is integrated using flow conservation and prediction formulas to compute the flow at next time. In the case of a micro cluster, it consists in changing the frequency of generation of vehicles and their speed based on data from the sensors, and measuring the number of vehicles passing the sensors and their speed, for a given time interval to deduct the sensor data.

 \subsubsection*{Aggregation and disaggregation of traffic {variables}}
 
When running the simulation, an agent manages the decomposition of the cluster network. This agent encapsulates rules, determining whether the model used in a cluster must be changed (for example from micro to macro) and determining whether the area of a cluster must be enlarged or reduced according to the needs of the simulation (see the previous section). Switching from micro to macro is straightforward: the mean speed and flow inside the cluster can be directly computed from sensors. On the other hand, switching from macro to micro requires to perform a local warm-up so the speed, density and flow of vehicles can be accurate.
 
\subsubsection*{Formal multi-level agent-based meta-model}

Regular multi-agent based simulation meta-models lack the structure to manage 
such hybrid approaches: their representation of the agents, the 
environment and the temporal dynamics of the system is designed to support a single 
viewpoint.
Multi-level agent-based modeling is an interesting approach to simulate such systems. Indeed it offers a large range of techniques to dynamically adapt  the level of detail of simulations, couple heterogenous models or detect and reify emergent phenomena~\citep{Gaud:2008,Gil-Quijano:2012,Picault:2011}.

Many meta-models and simulation engines dedicated to multi-level agent-based modeling have been proposed in the literature. See \textit{e.g.} \citet{Morvan:2013} for a complete review. Managing multiple viewpoints on the same phenomenon induces the use of heterogeneous time models, thus 
raising issues related to \emph{time} and \emph{consistency}. However, simulation bias can be controlled  by rules that constraint perceptions, influence production and reaction computation, according to causality and coherence principles~\citep{Morvan:2011}.
To deal with this issue, we developed a generic approach called \textsc{Similar} to design simulations (see \citet{Morvan:2014b} for more  detailed  information about the main principles of \textsc{Similar}). Indeed, this approach relies on a multi-level, influence-reaction and agent-based knowledge representation, more fitting to multiple viewpoints~\citep{Ferber:1996,Michel:2007,Soyez:2013}.

\section*{The \textsc{Jam-free} simulator}

\subsection*{Architecture}

\begin{figure}[ht]
	\centering
	\includegraphics[width=\linewidth]{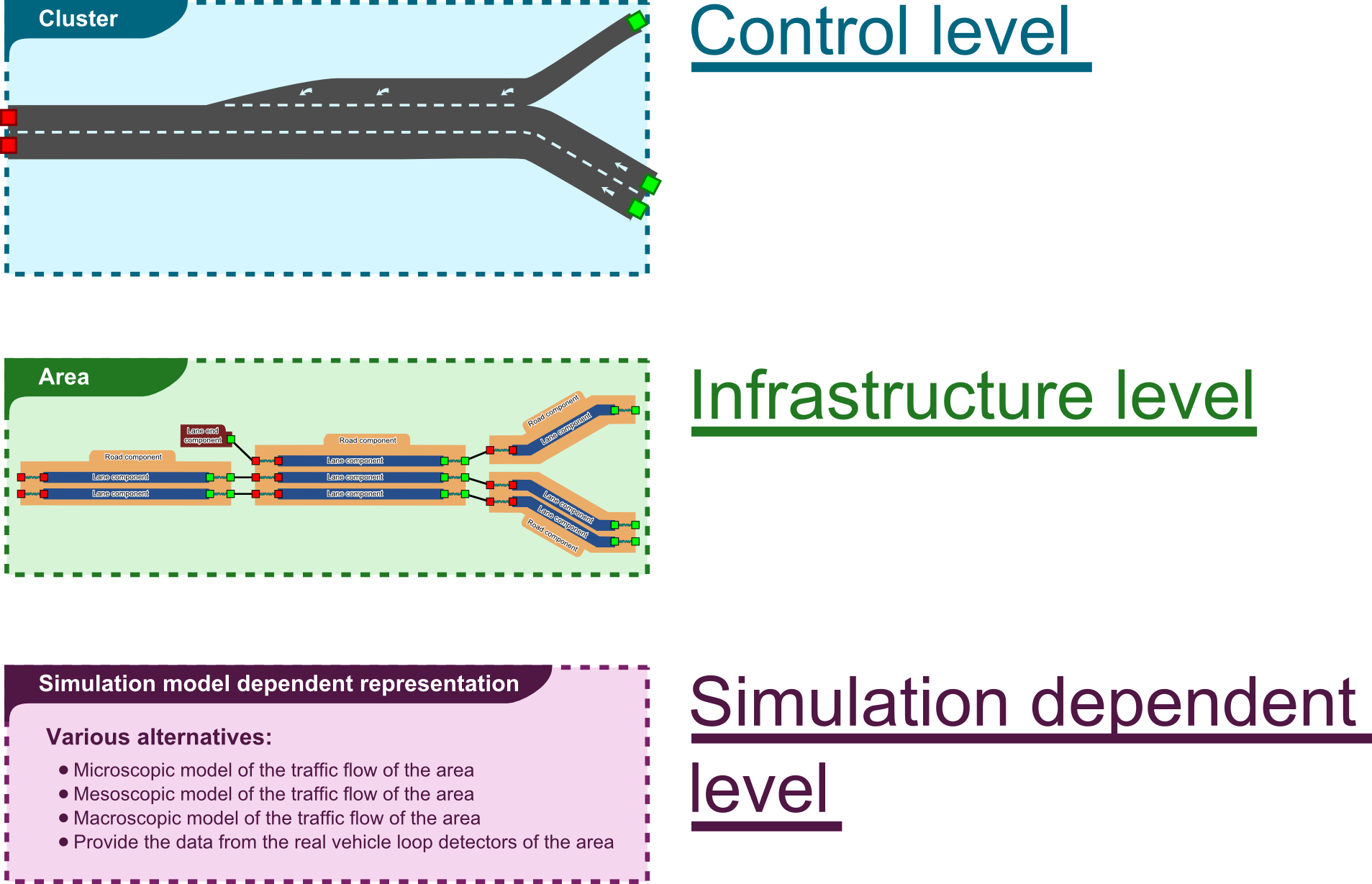}
	\caption{Multi-level architecture of \textsc{Jam-free}}
	\label{img:multi-level  architecture}
\end{figure}

\textsc{Jam-free} is based on a multi-level architecture~(see e.g., \ figure~\ref{img:multi-level  architecture}):
\begin{itemize}
\item the \textit{infrastructure} level models the components of the road network; the naming and identification of road network components follow the S\'etra (Service d'\'etudes sur les transports, les routes et leurs am'enagements, is a French technical service of the  Minist\`ere de l'\'Ecologie, du D\'eveloppement Durable et de l'\'Energie dedicated to transportation issues) specification~\citep{Setra:2010}.
\item the \textit{control} level is the core of the dynamic hybrid approach: it manages the dynamic decomposition of the network into clusters.
\item the \textit{simulation dependent} levels model the traffic dynamic using a dedicated representation; by default, three traffic levels are considered: microscopic, macroscopic and data-driven (\textit{i.e.}, using real data).
\end{itemize}

\subsection*{Models used in \textsc{Jam-free}}

\subsubsection*{Road network structure}

We consider that the road network can be divided dynamically into subsets called clusters. The traffic is simulated on each cluster using various heterogeneous models depending on the situation: either a microscopic model or a macroscopic model. For computation time efficiency reasons, we choose not to model the road network at a physical level. Indeed, such a model requires the vehicles to interpret a large continuous zone of asphalt with blank lines as separated lanes. Such computations are unnecessary complex, since in our use case (France) rational drivers usually never drive over the blank lines separating the lanes if they are not overtaking. To keep the model simple, the road network is only modeled at a semantic level.

The road network is modeled as a set of interconnected roads. Each road contains a set of lanes (usually from 1 to 5). Roads also contain vertical signs (\textit{e.g.} stop sign, speed limit sign) that are bond either to all the lanes or to specific lanes (\textit{e.g.}  extraction lane in the highway, slow vehicles lane). In our model, we support a specific subset of the french vertical signs.
The connection points of the roads are called nodes. We distinguish different types of nodes:
\begin{itemize}
\item Crossroads nodes
\item Roundabout nodes
\item Highway insertion node
\item Highway extraction node
\end{itemize}

\subsubsection*{Microscopic level models}

The behavior of a vehicle can be seen as the sum of three different parts:
\begin{itemize}
\item A navigation behavior when the current lane of the vehicle does not lead to the desired destination of the vehicle. This behavior consists in changing the lane of the vehicle until the current lane of the vehicle leads to the desired destination.
\item An overtaking behavior when the vehicle either wants to increase its speed by changing its lane, or when the vehicle has to swerve because a faster vehicle is tailing it.
\item A behavior when no lane change is required (acceleration model).
\end{itemize}
\textsc{Jam-free} implements the highly used Intelligent-Driver Model (\textsc{Idm}) to model the acceleration behavior of vehicles~\citep{Kesting:2010}.

\textsc{Idm} is a microscopic traffic flow model, \textit{i.e.}, each vehicle-driver combination constitutes an active "particle" in the simulation. Such models characterize the traffic state at any given time by the positions and speeds of all simulated vehicles. In case of multi-lane traffic, the lane index complements the state description. More specifically, \textsc{Idm} is a car-following model. In such models, the decision of any driver to accelerate or to brake depends only on his or her own speed, and on the position and speed of the "leading vehicle" immediately ahead. The model structure of  \textsc{Idm} can be described as follows:
\begin{itemize}
\item the influencing factors (model input) are the own speed v, the bumper-to-bumper gap s to the leading vehicle, and the relative speed (speed difference) of the two vehicles (positive when approaching),
\item the model output is the acceleration chosen by the driver for this situation,
\item the model parameters describe the driving style, \textit{i.e.}, whether the simulated driver drives slow or fast, careful or reckless.
\end{itemize}

Lane changes takes place, if:
\begin{itemize}
\item the potential new target lane is more attractive, i.e., the "incentive criterion" is satisfied,
\item and the change can be performed safely, i.e., the "safety criterion" is satisfied.
\end{itemize}
\textsc{Jam-free} implements the lane changing model \textsc{Mobil}~\citep{Kesting:2007}.

Chosen acceleration and lane changing models are presented in detail in the \textsc{Jam-free} documentation.

\subsubsection*{Macroscopic level model}

Any macroscopic traffic flow model can be used in \textsc{Jam-free} since it defines generic connectors to convert micro/macro representations. By default, a generic implementation of  the \textsc{Metanet} model is provided~\citep{Messner:1990}. {\textsc{Metanet} is based on a second order macroscopic model and uses the following traffic flow equations:
\begin{equation}\label{eq:cons}
\rho_i(k+1)=\rho_i(k)+\frac{T_s}{L_i}\left[q_{i-1}(k)-q_i(k)\right]
\end{equation}
where, $\rho_i$, defines the traffic density in ($veh/km/lane$). $L_i$, $k$ and $T_s$ represent, the segment length (km), the simulation step and its duration, respectively.  $q_i$ in ($vehicles/h$) defines the traffic flow in the segment $i$:
\begin{equation}\label{relation1}
q_i=\rho_iv_i
\end{equation}
The speed $v_i$ in (km/h) represents the velocity of vehicles in segment $i$ at time  $kT_s$. \begin{equation}\label{eq_vitesse}
v_i=V_e\left(\rho_i\right)
\end{equation}
where $V_e\left(\rho_i\right)$ is the static  speed-density characteristic called a fundamental diagram.  The momentum equation is obtained by introducing two small constant parameters in \eqref{eq_vitesse}. Such equation reads:
\begin{equation*}\label{eq_m}
v_{i}(k+1)=\underbrace{\frac{T_s}{\tau}[V_e(\rho_{i})-v_{i}(k)]}_{relaxation}+\underbrace{\frac{T_s\eta}{L_i}v_{i}(k)[v_{i-1}(k)-v_{i+1}(k)]}_{convection}\end{equation*}
\begin{equation}\label{eq_m2}
-\underbrace{\frac{T_s\nu}{\tau L_i}\frac{\rho_{i+1}(k)-\rho_{i}(k)}{\rho_{i}(k) +\kappa}}_{anticipation}
\end{equation}
The first term of equation \eqref{eq_m2} represents relaxation to the equilibrium. This is the most dominant term in the equation since the other terms  manifest their effects only when there are (occasional) fluctuations in the traffic speed and density upstream or downstream.  This relaxation term describes the fact that the drivers adjust their speeds to the equilibrium speed-density characteristic $V_e\left(\rho_i\right)$ according to the reaction time $\tau$.  
The second term, represents the convection, \textit{i.e.}, the influence of the upstream traffic mean speed. The third term, called "anticipation", translates the effect of drivers reacting to downstream traffic density  variations. Drivers tend to decelerate if the downstream density is higher and accelerate if it is lower. $\nu$, $\kappa$ are a model parameters.
}

\subsubsection*{Traffic generation}

In our model, each input point is a special connector that will generate traffic on the road it is attached to. Note that such a connector has to be created and put on each lane where the traffic appears. The creation of vehicles is managed by a traffic input point agent. 
Various implementation of this agent currently exist:
\begin{itemize}
\item "Flow-mass traffic input point" agents, generating the traffic flow using a flow-mass parameter
\item "Scripted traffic input point" agents, generating the traffic flow using user-defined events. The created vehicles and the creation dates are manually specified by the users.
\end{itemize}

\subsection*{A first experimentation}

\begin{figure}[ht]
	\includegraphics[width=\linewidth]{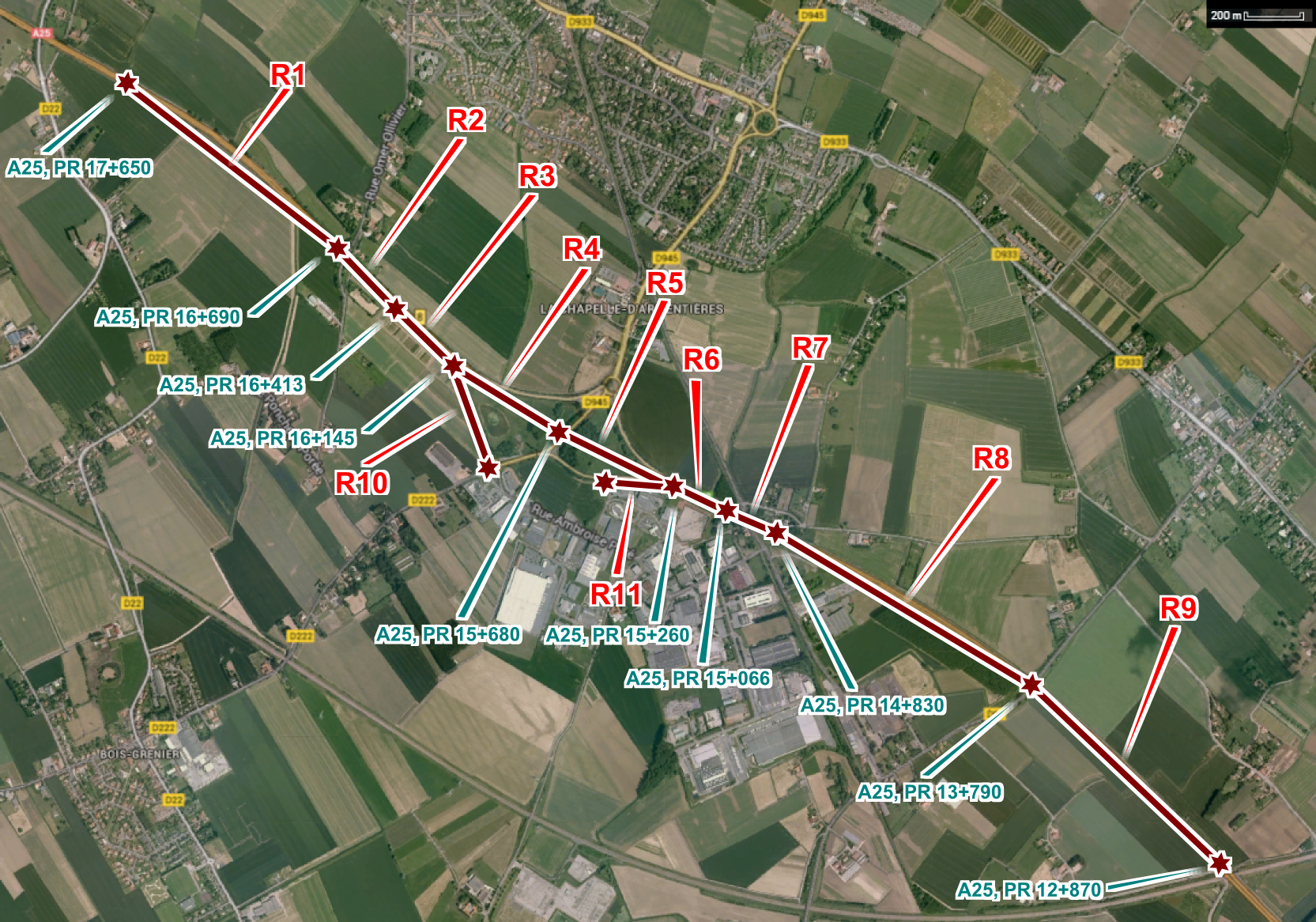}
	\caption{Description of the simulated network}
	\label{img:network overview}
\end{figure}

To validate the models implemented in \textsc{Jam-free}, as well as the flow (dis)aggregation algorithms, we performed a first hybrid simulation using real data from of the A25 highway in France. We simulate a 4780m portion of the highway, decomposed into 11 clusters~(see fig.~\ref{img:network overview}). The clusters containing insertion or extraction lanes are simulated with a microscopic model, the others with a macroscopic model.  

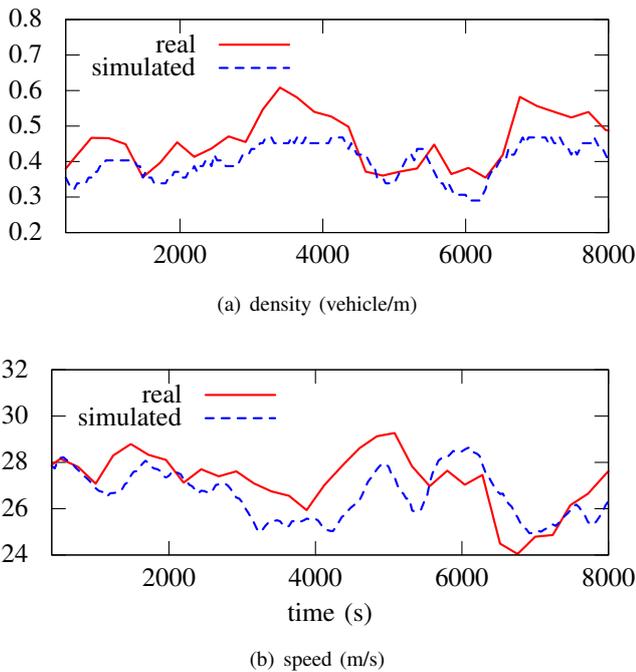
\begin{figure}[ht]
		\subfigure[density (vehicle/m)]{\begin{tikzpicture}[gnuplot]
%% generated with GNUPLOT 4.6p6 (Lua 5.2; terminal rev. 99, script rev. 100)
%% Dim  5 avr 11:57:37 2015
\path (0.000,0.000) rectangle (8.890,3.810);
\gpcolor{color=gp lt color border}
\gpsetlinetype{gp lt border}
\gpsetlinewidth{1.00}
\draw[gp path] (1.196,0.616)--(1.376,0.616);
\draw[gp path] (8.337,0.616)--(8.157,0.616);
\node[gp node right] at (1.012,0.616) { 0.2};
\draw[gp path] (1.196,1.087)--(1.376,1.087);
\draw[gp path] (8.337,1.087)--(8.157,1.087);
\node[gp node right] at (1.012,1.087) { 0.3};
\draw[gp path] (1.196,1.558)--(1.376,1.558);
\draw[gp path] (8.337,1.558)--(8.157,1.558);
\node[gp node right] at (1.012,1.558) { 0.4};
\draw[gp path] (1.196,2.028)--(1.376,2.028);
\draw[gp path] (8.337,2.028)--(8.157,2.028);
\node[gp node right] at (1.012,2.028) { 0.5};
\draw[gp path] (1.196,2.499)--(1.376,2.499);
\draw[gp path] (8.337,2.499)--(8.157,2.499);
\node[gp node right] at (1.012,2.499) { 0.6};
\draw[gp path] (1.196,2.970)--(1.376,2.970);
\draw[gp path] (8.337,2.970)--(8.157,2.970);
\node[gp node right] at (1.012,2.970) { 0.7};
\draw[gp path] (1.196,3.441)--(1.376,3.441);
\draw[gp path] (8.337,3.441)--(8.157,3.441);
\node[gp node right] at (1.012,3.441) { 0.8};
\draw[gp path] (2.699,0.616)--(2.699,0.796);
\draw[gp path] (2.699,3.441)--(2.699,3.261);
\node[gp node center] at (2.699,0.308) { 2000};
\draw[gp path] (4.579,0.616)--(4.579,0.796);
\draw[gp path] (4.579,3.441)--(4.579,3.261);
\node[gp node center] at (4.579,0.308) { 4000};
\draw[gp path] (6.458,0.616)--(6.458,0.796);
\draw[gp path] (6.458,3.441)--(6.458,3.261);
\node[gp node center] at (6.458,0.308) { 6000};
\draw[gp path] (8.337,0.616)--(8.337,0.796);
\draw[gp path] (8.337,3.441)--(8.337,3.261);
\node[gp node center] at (8.337,0.308) { 8000};
\draw[gp path] (1.196,3.441)--(1.196,0.616)--(8.337,0.616)--(8.337,3.441)--cycle;
\node[gp node right] at (3.036,3.107) {real};
\gpcolor{color=gp lt color 0}
\gpsetlinetype{gp lt plot 0}
\gpsetlinewidth{2.00}
\draw[gp path] (3.220,3.107)--(4.136,3.107);
\draw[gp path] (1.196,1.464)--(1.234,1.510)--(1.271,1.556)--(1.309,1.603)--(1.346,1.648)%
  --(1.384,1.692)--(1.422,1.737)--(1.459,1.782)--(1.497,1.827)--(1.534,1.872)--(1.572,1.871)%
  --(1.609,1.870)--(1.647,1.870)--(1.685,1.869)--(1.722,1.868)--(1.760,1.868)--(1.797,1.854)%
  --(1.835,1.841)--(1.873,1.827)--(1.910,1.814)--(1.948,1.801)--(1.985,1.787)--(2.023,1.714)%
  --(2.060,1.641)--(2.098,1.568)--(2.136,1.496)--(2.173,1.423)--(2.211,1.350)--(2.248,1.381)%
  --(2.286,1.413)--(2.324,1.445)--(2.361,1.477)--(2.399,1.508)--(2.436,1.540)--(2.474,1.585)%
  --(2.511,1.631)--(2.549,1.676)--(2.587,1.722)--(2.624,1.767)--(2.662,1.813)--(2.699,1.781)%
  --(2.737,1.749)--(2.775,1.717)--(2.812,1.685)--(2.850,1.652)--(2.887,1.620)--(2.925,1.639)%
  --(2.962,1.657)--(3.000,1.675)--(3.038,1.694)--(3.075,1.712)--(3.113,1.730)--(3.150,1.757)%
  --(3.188,1.784)--(3.226,1.811)--(3.263,1.838)--(3.301,1.864)--(3.338,1.891)--(3.376,1.879)%
  --(3.413,1.866)--(3.451,1.854)--(3.489,1.841)--(3.526,1.829)--(3.564,1.817)--(3.601,1.889)%
  --(3.639,1.960)--(3.677,2.032)--(3.714,2.104)--(3.752,2.176)--(3.789,2.248)--(3.827,2.297)%
  --(3.864,2.345)--(3.902,2.393)--(3.940,2.442)--(3.977,2.490)--(4.015,2.539)--(4.052,2.517)%
  --(4.090,2.495)--(4.128,2.473)--(4.165,2.451)--(4.203,2.429)--(4.240,2.407)--(4.278,2.375)%
  --(4.315,2.344)--(4.353,2.312)--(4.391,2.280)--(4.428,2.249)--(4.466,2.217)--(4.503,2.206)%
  --(4.541,2.196)--(4.579,2.185)--(4.616,2.175)--(4.654,2.165)--(4.691,2.154)--(4.729,2.131)%
  --(4.767,2.109)--(4.804,2.086)--(4.842,2.064)--(4.879,2.041)--(4.917,2.019)--(4.954,1.920)%
  --(4.992,1.821)--(5.030,1.721)--(5.067,1.622)--(5.105,1.523)--(5.142,1.424)--(5.180,1.415)%
  --(5.218,1.407)--(5.255,1.398)--(5.293,1.389)--(5.330,1.380)--(5.368,1.371)--(5.405,1.380)%
  --(5.443,1.389)--(5.481,1.398)--(5.518,1.407)--(5.556,1.415)--(5.593,1.424)--(5.631,1.431)%
  --(5.669,1.438)--(5.706,1.445)--(5.744,1.452)--(5.781,1.459)--(5.819,1.465)--(5.856,1.518)%
  --(5.894,1.571)--(5.932,1.624)--(5.969,1.677)--(6.007,1.730)--(6.044,1.783)--(6.082,1.718)%
  --(6.120,1.652)--(6.157,1.587)--(6.195,1.522)--(6.232,1.456)--(6.270,1.391)--(6.307,1.405)%
  --(6.345,1.418)--(6.383,1.432)--(6.420,1.446)--(6.458,1.460)--(6.495,1.473)--(6.533,1.452)%
  --(6.571,1.430)--(6.608,1.409)--(6.646,1.387)--(6.683,1.365)--(6.721,1.344)--(6.758,1.394)%
  --(6.796,1.444)--(6.834,1.494)--(6.871,1.544)--(6.909,1.594)--(6.946,1.644)--(6.984,1.772)%
  --(7.022,1.901)--(7.059,2.029)--(7.097,2.158)--(7.134,2.286)--(7.172,2.415)--(7.209,2.395)%
  --(7.247,2.374)--(7.285,2.354)--(7.322,2.334)--(7.360,2.314)--(7.397,2.293)--(7.435,2.281)%
  --(7.473,2.268)--(7.510,2.255)--(7.548,2.242)--(7.585,2.230)--(7.623,2.217)--(7.660,2.204)%
  --(7.698,2.192)--(7.736,2.180)--(7.773,2.167)--(7.811,2.155)--(7.848,2.142)--(7.886,2.154)%
  --(7.924,2.166)--(7.961,2.179)--(7.999,2.191)--(8.036,2.203)--(8.074,2.215)--(8.111,2.175)%
  --(8.149,2.135)--(8.187,2.095)--(8.224,2.055)--(8.262,2.015)--(8.299,1.976)--(8.337,1.968);
\gpcolor{color=gp lt color border}
\node[gp node right] at (3.036,2.799) {simulated};
\gpcolor{color=gp lt color 2}
\gpsetlinetype{gp lt plot 1}
\draw[gp path] (3.220,2.799)--(4.136,2.799);
\draw[gp path] (1.196,1.345)--(1.234,1.269)--(1.271,1.193)--(1.309,1.193)--(1.346,1.269)%
  --(1.384,1.269)--(1.422,1.269)--(1.459,1.269)--(1.497,1.345)--(1.534,1.345)--(1.572,1.421)%
  --(1.609,1.421)--(1.647,1.497)--(1.685,1.497)--(1.722,1.573)--(1.760,1.573)--(1.797,1.573)%
  --(1.835,1.573)--(1.873,1.573)--(1.910,1.573)--(1.948,1.573)--(1.985,1.573)--(2.023,1.573)%
  --(2.060,1.573)--(2.098,1.573)--(2.136,1.497)--(2.173,1.497)--(2.211,1.345)--(2.248,1.421)%
  --(2.286,1.345)--(2.324,1.345)--(2.361,1.269)--(2.399,1.269)--(2.436,1.269)--(2.474,1.269)%
  --(2.511,1.269)--(2.549,1.345)--(2.587,1.345)--(2.624,1.421)--(2.662,1.421)--(2.699,1.421)%
  --(2.737,1.345)--(2.775,1.345)--(2.812,1.421)--(2.850,1.421)--(2.887,1.497)--(2.925,1.421)%
  --(2.962,1.497)--(3.000,1.497)--(3.038,1.573)--(3.075,1.573)--(3.113,1.497)--(3.150,1.649)%
  --(3.188,1.573)--(3.226,1.573)--(3.263,1.573)--(3.301,1.497)--(3.338,1.497)--(3.376,1.497)%
  --(3.413,1.497)--(3.451,1.497)--(3.489,1.573)--(3.526,1.649)--(3.564,1.649)--(3.601,1.649)%
  --(3.639,1.725)--(3.677,1.725)--(3.714,1.725)--(3.752,1.801)--(3.789,1.801)--(3.827,1.877)%
  --(3.864,1.877)--(3.902,1.877)--(3.940,1.801)--(3.977,1.801)--(4.015,1.801)--(4.052,1.801)%
  --(4.090,1.801)--(4.128,1.801)--(4.165,1.877)--(4.203,1.801)--(4.240,1.877)--(4.278,1.801)%
  --(4.315,1.801)--(4.353,1.877)--(4.391,1.801)--(4.428,1.801)--(4.466,1.801)--(4.503,1.801)%
  --(4.541,1.801)--(4.579,1.801)--(4.616,1.801)--(4.654,1.801)--(4.691,1.877)--(4.729,1.877)%
  --(4.767,1.877)--(4.804,1.877)--(4.842,1.877)--(4.879,1.801)--(4.917,1.725)--(4.954,1.801)%
  --(4.992,1.725)--(5.030,1.649)--(5.067,1.649)--(5.105,1.649)--(5.142,1.649)--(5.180,1.573)%
  --(5.218,1.497)--(5.255,1.497)--(5.293,1.345)--(5.330,1.345)--(5.368,1.345)--(5.405,1.269)%
  --(5.443,1.269)--(5.481,1.269)--(5.518,1.269)--(5.556,1.345)--(5.593,1.421)--(5.631,1.497)%
  --(5.669,1.573)--(5.706,1.649)--(5.744,1.649)--(5.781,1.573)--(5.819,1.725)--(5.856,1.725)%
  --(5.894,1.725)--(5.932,1.649)--(5.969,1.573)--(6.007,1.497)--(6.044,1.421)--(6.082,1.345)%
  --(6.120,1.269)--(6.157,1.269)--(6.195,1.193)--(6.232,1.269)--(6.270,1.193)--(6.307,1.117)%
  --(6.345,1.117)--(6.383,1.117)--(6.420,1.117)--(6.458,1.117)--(6.495,1.041)--(6.533,1.041)%
  --(6.571,1.041)--(6.608,1.041)--(6.646,1.041)--(6.683,1.117)--(6.721,1.193)--(6.758,1.345)%
  --(6.796,1.421)--(6.834,1.421)--(6.871,1.497)--(6.909,1.573)--(6.946,1.573)--(6.984,1.573)%
  --(7.022,1.725)--(7.059,1.649)--(7.097,1.649)--(7.134,1.725)--(7.172,1.801)--(7.209,1.877)%
  --(7.247,1.801)--(7.285,1.877)--(7.322,1.877)--(7.360,1.877)--(7.397,1.877)--(7.435,1.877)%
  --(7.473,1.877)--(7.510,1.877)--(7.548,1.801)--(7.585,1.877)--(7.623,1.801)--(7.660,1.801)%
  --(7.698,1.877)--(7.736,1.801)--(7.773,1.725)--(7.811,1.725)--(7.848,1.649)--(7.886,1.725)%
  --(7.924,1.649)--(7.961,1.725)--(7.999,1.801)--(8.036,1.801)--(8.074,1.801)--(8.111,1.801)%
  --(8.149,1.877)--(8.187,1.801)--(8.224,1.725)--(8.262,1.725)--(8.299,1.649)--(8.337,1.573);
\gpcolor{color=gp lt color border}
\gpsetlinetype{gp lt border}
\gpsetlinewidth{1.00}
\draw[gp path] (1.196,3.441)--(1.196,0.616)--(8.337,0.616)--(8.337,3.441)--cycle;
%% coordinates of the plot area
\gpdefrectangularnode{gp plot 1}{\pgfpoint{1.196cm}{0.616cm}}{\pgfpoint{8.337cm}{3.441cm}}
\end{tikzpicture}
%% gnuplot variables
}
		\subfigure[speed (m/s)]{\begin{tikzpicture}[gnuplot]
%% generated with GNUPLOT 4.6p6 (Lua 5.2; terminal rev. 99, script rev. 100)
%% Dim  5 avr 11:57:37 2015
\path (0.000,0.000) rectangle (8.890,3.810);
\gpcolor{color=gp lt color border}
\gpsetlinetype{gp lt border}
\gpsetlinewidth{1.00}
\draw[gp path] (1.012,0.985)--(1.192,0.985);
\draw[gp path] (8.337,0.985)--(8.157,0.985);
\node[gp node right] at (0.828,0.985) { 24};
\draw[gp path] (1.012,1.599)--(1.192,1.599);
\draw[gp path] (8.337,1.599)--(8.157,1.599);
\node[gp node right] at (0.828,1.599) { 26};
\draw[gp path] (1.012,2.213)--(1.192,2.213);
\draw[gp path] (8.337,2.213)--(8.157,2.213);
\node[gp node right] at (0.828,2.213) { 28};
\draw[gp path] (1.012,2.827)--(1.192,2.827);
\draw[gp path] (8.337,2.827)--(8.157,2.827);
\node[gp node right] at (0.828,2.827) { 30};
\draw[gp path] (1.012,3.441)--(1.192,3.441);
\draw[gp path] (8.337,3.441)--(8.157,3.441);
\node[gp node right] at (0.828,3.441) { 32};
\draw[gp path] (2.554,0.985)--(2.554,1.165);
\draw[gp path] (2.554,3.441)--(2.554,3.261);
\node[gp node center] at (2.554,0.677) { 2000};
\draw[gp path] (4.482,0.985)--(4.482,1.165);
\draw[gp path] (4.482,3.441)--(4.482,3.261);
\node[gp node center] at (4.482,0.677) { 4000};
\draw[gp path] (6.409,0.985)--(6.409,1.165);
\draw[gp path] (6.409,3.441)--(6.409,3.261);
\node[gp node center] at (6.409,0.677) { 6000};
\draw[gp path] (8.337,0.985)--(8.337,1.165);
\draw[gp path] (8.337,3.441)--(8.337,3.261);
\node[gp node center] at (8.337,0.677) { 8000};
\draw[gp path] (1.012,3.441)--(1.012,0.985)--(8.337,0.985)--(8.337,3.441)--cycle;
\node[gp node center] at (4.674,0.215) {time (s)};
\node[gp node right] at (2.852,3.107) {real};
\gpcolor{color=gp lt color 0}
\gpsetlinetype{gp lt plot 0}
\gpsetlinewidth{2.00}
\draw[gp path] (3.036,3.107)--(3.952,3.107);
\draw[gp path] (1.012,2.181)--(1.051,2.208)--(1.089,2.235)--(1.128,2.261)--(1.166,2.243)%
  --(1.205,2.225)--(1.243,2.207)--(1.282,2.189)--(1.320,2.171)--(1.359,2.153)--(1.398,2.116)%
  --(1.436,2.079)--(1.475,2.042)--(1.513,2.005)--(1.552,1.968)--(1.590,1.931)--(1.629,1.993)%
  --(1.667,2.056)--(1.706,2.118)--(1.745,2.180)--(1.783,2.243)--(1.822,2.305)--(1.860,2.330)%
  --(1.899,2.355)--(1.937,2.381)--(1.976,2.406)--(2.014,2.431)--(2.053,2.456)--(2.091,2.432)%
  --(2.130,2.409)--(2.169,2.385)--(2.207,2.361)--(2.246,2.337)--(2.284,2.313)--(2.323,2.302)%
  --(2.361,2.291)--(2.400,2.279)--(2.438,2.268)--(2.477,2.257)--(2.516,2.245)--(2.554,2.195)%
  --(2.593,2.145)--(2.631,2.095)--(2.670,2.045)--(2.708,1.994)--(2.747,1.944)--(2.785,1.974)%
  --(2.824,2.003)--(2.863,2.033)--(2.901,2.062)--(2.940,2.092)--(2.978,2.122)--(3.017,2.106)%
  --(3.055,2.090)--(3.094,2.075)--(3.132,2.059)--(3.171,2.044)--(3.210,2.028)--(3.248,2.039)%
  --(3.287,2.050)--(3.325,2.061)--(3.364,2.072)--(3.402,2.083)--(3.441,2.094)--(3.479,2.067)%
  --(3.518,2.041)--(3.556,2.014)--(3.595,1.987)--(3.634,1.961)--(3.672,1.934)--(3.711,1.917)%
  --(3.749,1.899)--(3.788,1.881)--(3.826,1.864)--(3.865,1.846)--(3.903,1.828)--(3.942,1.819)%
  --(3.981,1.809)--(4.019,1.799)--(4.058,1.789)--(4.096,1.780)--(4.135,1.770)--(4.173,1.738)%
  --(4.212,1.706)--(4.250,1.675)--(4.289,1.643)--(4.328,1.611)--(4.366,1.580)--(4.405,1.635)%
  --(4.443,1.690)--(4.482,1.745)--(4.520,1.799)--(4.559,1.854)--(4.597,1.909)--(4.636,1.952)%
  --(4.675,1.994)--(4.713,2.036)--(4.752,2.078)--(4.790,2.120)--(4.829,2.162)--(4.867,2.202)%
  --(4.906,2.242)--(4.944,2.281)--(4.983,2.321)--(5.021,2.360)--(5.060,2.400)--(5.099,2.427)%
  --(5.137,2.453)--(5.176,2.480)--(5.214,2.506)--(5.253,2.533)--(5.291,2.559)--(5.330,2.566)%
  --(5.368,2.574)--(5.407,2.581)--(5.446,2.588)--(5.484,2.595)--(5.523,2.602)--(5.561,2.528)%
  --(5.600,2.454)--(5.638,2.380)--(5.677,2.306)--(5.715,2.232)--(5.754,2.158)--(5.793,2.115)%
  --(5.831,2.071)--(5.870,2.028)--(5.908,1.985)--(5.947,1.941)--(5.985,1.898)--(6.024,1.932)%
  --(6.062,1.966)--(6.101,2.001)--(6.140,2.035)--(6.178,2.069)--(6.217,2.103)--(6.255,2.072)%
  --(6.294,2.041)--(6.332,2.010)--(6.371,1.979)--(6.409,1.948)--(6.448,1.917)--(6.486,1.938)%
  --(6.525,1.960)--(6.564,1.982)--(6.602,2.004)--(6.641,2.025)--(6.679,2.047)--(6.718,1.895)%
  --(6.756,1.743)--(6.795,1.592)--(6.833,1.440)--(6.872,1.288)--(6.911,1.136)--(6.949,1.113)%
  --(6.988,1.090)--(7.026,1.067)--(7.065,1.044)--(7.103,1.021)--(7.142,0.998)--(7.180,1.036)%
  --(7.219,1.074)--(7.258,1.112)--(7.296,1.151)--(7.335,1.189)--(7.373,1.227)--(7.412,1.231)%
  --(7.450,1.235)--(7.489,1.239)--(7.527,1.242)--(7.566,1.246)--(7.605,1.250)--(7.643,1.316)%
  --(7.682,1.382)--(7.720,1.448)--(7.759,1.514)--(7.797,1.580)--(7.836,1.646)--(7.874,1.671)%
  --(7.913,1.696)--(7.951,1.721)--(7.990,1.747)--(8.029,1.772)--(8.067,1.797)--(8.106,1.840)%
  --(8.144,1.883)--(8.183,1.926)--(8.221,1.969)--(8.260,2.012)--(8.298,2.055)--(8.337,2.096);
\gpcolor{color=gp lt color border}
\node[gp node right] at (2.852,2.799) {simulated};
\gpcolor{color=gp lt color 2}
\gpsetlinetype{gp lt plot 1}
\draw[gp path] (3.036,2.799)--(3.952,2.799);
\draw[gp path] (1.012,2.163)--(1.051,2.132)--(1.089,2.230)--(1.128,2.272)--(1.166,2.280)%
  --(1.205,2.246)--(1.243,2.223)--(1.282,2.199)--(1.320,2.149)--(1.359,2.112)--(1.398,2.071)%
  --(1.436,2.035)--(1.475,2.001)--(1.513,1.963)--(1.552,1.915)--(1.590,1.871)--(1.629,1.838)%
  --(1.667,1.824)--(1.706,1.816)--(1.745,1.765)--(1.783,1.804)--(1.822,1.807)--(1.860,1.812)%
  --(1.899,1.846)--(1.937,1.856)--(1.976,1.925)--(2.014,1.949)--(2.053,2.037)--(2.091,2.090)%
  --(2.130,2.097)--(2.169,2.152)--(2.207,2.196)--(2.246,2.235)--(2.284,2.203)--(2.323,2.155)%
  --(2.361,2.131)--(2.400,2.124)--(2.438,2.100)--(2.477,2.034)--(2.516,2.012)--(2.554,2.036)%
  --(2.593,2.084)--(2.631,2.103)--(2.670,2.073)--(2.708,2.056)--(2.747,1.989)--(2.785,1.962)%
  --(2.824,1.915)--(2.863,1.894)--(2.901,1.845)--(2.940,1.804)--(2.978,1.838)--(3.017,1.833)%
  --(3.055,1.818)--(3.094,1.805)--(3.132,1.846)--(3.171,1.913)--(3.210,1.922)--(3.248,1.926)%
  --(3.287,1.933)--(3.325,1.909)--(3.364,1.864)--(3.402,1.799)--(3.441,1.747)--(3.479,1.732)%
  --(3.518,1.684)--(3.556,1.608)--(3.595,1.549)--(3.634,1.481)--(3.672,1.415)--(3.711,1.330)%
  --(3.749,1.287)--(3.788,1.324)--(3.826,1.376)--(3.865,1.416)--(3.903,1.436)--(3.942,1.439)%
  --(3.981,1.445)--(4.019,1.430)--(4.058,1.377)--(4.096,1.356)--(4.135,1.370)--(4.173,1.432)%
  --(4.212,1.435)--(4.250,1.415)--(4.289,1.429)--(4.328,1.444)--(4.366,1.467)--(4.405,1.459)%
  --(4.443,1.457)--(4.482,1.464)--(4.520,1.439)--(4.559,1.382)--(4.597,1.345)--(4.636,1.316)%
  --(4.675,1.305)--(4.713,1.302)--(4.752,1.359)--(4.790,1.422)--(4.829,1.489)--(4.867,1.516)%
  --(4.906,1.568)--(4.944,1.610)--(4.983,1.641)--(5.021,1.674)--(5.060,1.709)--(5.099,1.786)%
  --(5.137,1.879)--(5.176,1.961)--(5.214,2.009)--(5.253,2.068)--(5.291,2.122)--(5.330,2.168)%
  --(5.368,2.196)--(5.407,2.191)--(5.446,2.152)--(5.484,2.047)--(5.523,2.018)--(5.561,1.908)%
  --(5.600,1.837)--(5.638,1.798)--(5.677,1.731)--(5.715,1.734)--(5.754,1.575)--(5.793,1.558)%
  --(5.831,1.572)--(5.870,1.586)--(5.908,1.682)--(5.947,1.751)--(5.985,1.942)--(6.024,2.073)%
  --(6.062,2.148)--(6.101,2.237)--(6.140,2.267)--(6.178,2.264)--(6.217,2.284)--(6.255,2.320)%
  --(6.294,2.365)--(6.332,2.347)--(6.371,2.343)--(6.409,2.359)--(6.448,2.384)--(6.486,2.404)%
  --(6.525,2.412)--(6.564,2.346)--(6.602,2.331)--(6.641,2.299)--(6.679,2.185)--(6.718,2.083)%
  --(6.756,2.032)--(6.795,1.958)--(6.833,1.892)--(6.872,1.834)--(6.911,1.794)--(6.949,1.799)%
  --(6.988,1.713)--(7.026,1.662)--(7.065,1.624)--(7.103,1.547)--(7.142,1.483)--(7.180,1.406)%
  --(7.219,1.358)--(7.258,1.325)--(7.296,1.274)--(7.335,1.277)--(7.373,1.288)--(7.412,1.305)%
  --(7.450,1.293)--(7.489,1.320)--(7.527,1.339)--(7.566,1.372)--(7.605,1.390)--(7.643,1.370)%
  --(7.682,1.409)--(7.720,1.436)--(7.759,1.501)--(7.797,1.533)--(7.836,1.578)--(7.874,1.630)%
  --(7.913,1.654)--(7.951,1.619)--(7.990,1.557)--(8.029,1.505)--(8.067,1.422)--(8.106,1.396)%
  --(8.144,1.388)--(8.183,1.438)--(8.221,1.495)--(8.260,1.557)--(8.298,1.643)--(8.337,1.698);
\gpcolor{color=gp lt color border}
\gpsetlinetype{gp lt border}
\gpsetlinewidth{1.00}
\draw[gp path] (1.012,3.441)--(1.012,0.985)--(8.337,0.985)--(8.337,3.441)--cycle;
%% coordinates of the plot area
\gpdefrectangularnode{gp plot 1}{\pgfpoint{1.012cm}{0.985cm}}{\pgfpoint{8.337cm}{3.441cm}}
\end{tikzpicture}
%% gnuplot variables
}
		\caption{Simulation results in cluster $R8$}
		\label{img:results}
\end{figure}

The obtained results permit to validate the performance of \textsc{Jam-free} as well as its microscopic and macroscopic models. We cannot  present them all in this paper for an evident question of space. The figure~\ref{img:results} shows some simulation results in cluster $R8$ (\textit{i.e.},  after (dis)aggregation processes), using typical model parameter values (\textit{i.e.},  without specific calibration).

\section*{Conclusion and perspectives}

In this paper a first step towards the implementation of a dynamic hybrid model was proposed. The main objectives of this step were to identify use-cases, propose generic solutions to the  dynamic hybrid simulation problem and study the core aspects of the multi-level simulation approach. These ideas have then been implemented in a simulator called \textsc{Jam-free}.
The first experiments using real data allow us to validate several issues addressed in the paper and the implemented models. Further works will focus on the introduction of several traffic models allowing to the practitioners multiple choices in function of their simulation objective and the validation of the identified use-cases. Moreover, a more accurate macro/micro switching algorithm is being developed using a sophisticated continuous warm-up process. In addition, some control strategies will be added to \textsc{Jam-free} in order to evaluate their impact on the global behavior of the networks.

\bibliographystyle{ecms}
\bibliography{../../Biblio}

\end{document}